\documentclass{emulateapj}
\begin{document}














\title{Low frequency QPOs and possible change in the accretion
geometry during the outbursts of Aquila X$-$1}


\author{Wenda Zhang and Wenfei Yu} 
\affil{Shanghai Astronomical Observatory and Key Laboratory for
Research in Galaxies and Cosmology, Chinese Academy of Sciences, 80
Nandan Road, Shanghai, 200030, China; wenfei@shao.ac.cn}







\begin{abstract}
We have studied the evolution of the Low Frequency Quasi-Periodic
Oscillations (LFQPOs) during the rising phase of seven outbursts of
the neutron star Soft X-ray Transient (SXT) Aql X$-$1 observed with
the \textit{Rossi X-ray Timing Explorer (RXTE)}. The frequency
correlation between the low frequency break and the LFQPO sampled on
the time scale of $\sim$2 days was seen. Except for the peculiar 2001
outburst, the frequency of the LFQPOs increased with time before the
hard-to-soft state transition up to a maximum $\nu_{max}$ at $\sim$31
Hz, a factor of $\sim$5 higher than those seen in black hole
transients such as GX 339$-$4, making the maximum QPO frequency a
likely indicator of the mass of the central compact object. The
characteristic frequencies increased by around ten percent per day in
the early rising phase and accelerated to nearly one hundred percent
per day since $\sim$2 days before the hard-to-soft state transition.
We examined the dependence of the frequency $\nu_{LF}$ on the source
flux $f$ and found an anti-correlation between the maximum frequency
of the LFQPOs and the corresponding X-ray luminosity of the
hard-to-soft transition (or outburst peak luminosity) among the
outbursts. We suggest that X-ray evaporation process can not be the
only mechanism that drives the variation of the inner disk radius if
either of the twin kHz QPO corresponds to the Keplerian frequency at
the truncation radius.

\end{abstract} 


\section{Introduction}
Low Mass X-ray Binaries (LMXBs) are binary systems that contain a
black hole (BH) or neutron star (NS) primary and low mass companion.
The primary accretes from the companion through an accretion disk
around the compact object which emits predominately in the X-ray band
at high mass accretion rates. Soft X-ray transients (SXTs) are
transient LMXBs. They spend most of the time in the quiescent state
at low luminosity, and occasionally turn into outbursts during which
their luminosity increases by several orders of magnitude. During
outbursts, SXTs show distinctive states with coupled spectral and
timing properties: in the Low-Hard State (LHS) the soft X-ray (e.g.,
0.3--10 keV) energy spectrum is dominated by a power-law spectral
component and the Fourier power spectrum at low frequency is
dominated by broad band noise component and quasi-periodic
oscillations (QPOs), and the variability amplitude is quite large at
a few tens percent level, while in the High-Soft State (HSS) the soft
X-ray energy spectrum is dominated by a thermal disk component and
the Fourier power spectrum is dominated by a power-law noise
component with little variability \citep[see,
e.g.,][]{belloni_evolution_2005,remillard_x-ray_2006}.

Systematic studies of the state transitions in SXTs during the rising
phase of transient outbursts suggest that the accretion process is
non-stationary: the luminosity of the hard-to-soft state transitions
can occur at quite different luminosity, spanning two orders of
magnitude which correlates with the rate-of-increase of the
luminosity as well as the peak luminosity of the corresponding
outbursts
\citep{yu_correlation_2004,yu2007,yu_state_2009,tang_rxte/asm_2011}.
This indicates that the mass accretion rate cannot be the only
parameter that determines the transition between the accretion states
and the rate-of-change of the mass accretion rate is one important
parameter which determines the regimes of the accretion states in SXT
outbursts.

The most popular picture for the X-ray energy spectra in the distinct
states of black hole and neutron star SXTs is the truncated disk
picture, such as the model proposed in
\citet{esin_advection-dominated_1997}. In this picture the change of
the energy spectrum is associated with the change of the accretion
geometry. A geometrically-thin but optically-thick disk is truncated
at the truncation radius. The region within the radius is filled with
a hot flow such as the radiatively-inefficient accretion flow. In the
quiescent and the hard state, the truncation radius is quite large.
During the rising phase of transient outbursts the truncation radius
moves inward with increasing mass accretion rate, and reaches the
innermost stable orbit during the soft state. The detailed process
that truncates the disk is not clear, and proposals for the physical
mechanism include magnetosphere-disk interaction \citep[][for neutron
star systems only]{ghosh_accretion_1977} and disk evaporation
\citep[e.g.,][]{meyer_accretion_1994,liu_evaporation_1999,meyer_re-condensation_2007}.

It was suggested that timing features, especially the quasi-periodic
oscillations (QPOs) shown in the power spectra, might be the probe of
the accretion geometry and its change in X-ray binaries several
decades ago
\citep[e.g.,][]{klis85a,klis85b,miller_sonic-point_1998,stella_lense-thirring_1998,osherovich_kilohertz_1999,ingram2009}.
The variation in the characteristic frequencies of the QPOs or noise
components may reflect the change in the inner most edge of the
accretion disk or the size of the corona. Low Frequency
Quasi-Periodic Oscillations (LFQPOs) are the most obvious feature in
the Fourier power density spectra (PDS) of black hole SXTs in the
hard and the intermediate state. For black hole transients, the
LFQPOs can be classified into A, B and C-type QPOs
\citep{wijnands_complex_1999,remillard_characterizing_2002}. Among
them, the C-type QPO appears in the hard and hard-intermediate states
and its characteristic frequency appears to correlate with the
luminosity. There are similar QPOs and broad noise components in
neutron star LMXBs. For example, \citet{yu_hard_2003} showed that the
1--20 Hz LFQPOs and broad noise components in the neutron star
transient Aql X$-$1 are associated with the LFQPOs and broad noise
components commonly seen in black hole transients during outbursts.

Aql X$-$1 is a well-known neutron star transient with very frequent
outbursts
\citep[e.g.,][]{priedhorsky_long-term_1984,kitamoto_unstable_1993,simon_recurrence_2002}.
Its neutron star nature was revealed by the detection of type-I X-ray
bursts from the source
\citep[e.g.,][]{koyama_discovery_1981,czerny_persistent_1987}. Its
neutron star spin was detected as well
\citep{zhang98,casella_discovery_2008}. Based on its spectral and
temporal properties \citep{reig_correlated_2000}, Aql X$-$1 has been
classified as an ``Atoll'' source, of which great similarities of
both timing and spectral properties to those black hole X-ray
binaries have been seen. This makes Aql X$-$1 an ideal source for the
study of the evolution of the characteristic timescales in the
accretion flow during the rising phase of their outbursts, during
which drastic changes in the accretion geometry are expected. In this
$Paper$ we study the evolution of the LFQPOs during the rising phase
of seven outbursts of the neutron star SXT Aql X$-$1 with the
observations performed with the \textit{RXTE}. 





\section{Observations and Data Reduction}

We made use of the \textit{RXTE}/ASM and the MAXI long-term X-ray
monitoring observations of Aquila X$-$1 to identify X-ray outbursts
in this study. \textit{RXTE}/PCA pointed observations were used to
study X-ray timing and spectral properties. We have analysed a total
of 611 \textit{RXTE} observations of Aquila X$-$1 performed from 1996
to 2011, starting from the observation 10072-06-01-00 to the
observation 96440-01-10-05.

\subsection{Long-term Light Curves and Evolution of the Hardness Ratio}

Figure 1 shows the long-term X-ray light curve of Aquila X-1. We
converted the \textit{RXTE}/ASM 2--12 keV flux in unit of Crab by
dividing the ASM 2-12 keV count rate by the mean Crab count rate of
75.6 cts/s. Similarly, for the later period when \textit{RXTE}
stopped operation, we obtained the source flux in 2--10 keV by
combining the 2--4 keV and the 4--10 keV MAXI data which are publicly
available.

Figure 2 shows the X-ray light curve and the X-ray hardness ratio
seen with the pointed observations of the \textit{RXTE}/PCA. We
extracted source light curves from the PCA standard
products\footnote{http://heasarc.nasa.gov/docs/xte/recipes/stdprod\_guide.html}
which provide data with a time resolution of 16 seconds from the STD
2 PCA data. The light curves in the PCA standard products have five
energy bands, namely 2--9, 2--4, 4--9, 9--20 and 20--60 keV. We
identified the time intervals that were associated with Type-I bursts
following the screening method mentioned in
\citet{klein-wolt_identification_2008} and excluded these time
intervals. Then we calculated the mean count rates per PCU in the
five energy bands, and the hardness ratio (HR), which was defined as
the count rate ratio between the 9--20 keV and the 2--9 keV in each
observation.

\subsection{Timing Analysis}

We made use of the PCA data with a time resolution better than 0.25
ms and combined all available data in different energy bands whenever
possible. We extracted the data and rebinned the light curves into
1/4096 s resolution and then generated the Fourier power spectrum for
every 128 s segment. In the analysis of each power spectrum, we took
the average power above 1800 Hz to estimate the white noise level,
which was then subtracted. We fit each power spectrum with a model
composed of several Lorentzians in which zero-centered Lorentzian
represents the band-limited noise components and narrow Lorentzians
represents the QPOs following the previous approach
\citep{belloni_unified_2002}. A power-law component was included to
fit the Very Low Frequency Noises (VLFNs) when necessary. These
model fits were performed using a custom package making use of the
MPFIT
package\footnote{\url{http://cow.physics.wisc.edu/~craigm/idl/}}.
The best-fit model was achieved by minimizing the $\chi^2$, and the
uncertainty was obtained from the distribution of $\chi^2$.
Throughout this paper the error quoted for the power spectral fit
represents the 1$\sigma$ uncertainty.

\subsection{Spectral Analysis}

In order to estimate the X-ray flux of the source, we fit the 3--20
keV Standard 2 PCA energy spectra of the observations when the source
was in the hard state during the rising phase of each outburst. The
X-ray spectral model we used consisted of four components, namely a
power-law component representing the emission from the hot accretion
flow, a blackbody component for the emission from the neutron star
surface, a Gaussian line component with its central energy fixed at
6.5 keV representing a broad Iron emission line, and the absorption
by Galactic neutral hydrogen. We fixed the column density of the
neutral hydrogen to $3.4\times10^{21}~\rm atoms~cm^{-2}$, according
to \citet{dickey_h_1990}. The 2--20 keV flux was then obtained from
the spectral fits in XSPEC 12.8.2. 

\section{Results} 
\subsection{Outburst Sample}

\begin{figure}
\plotone{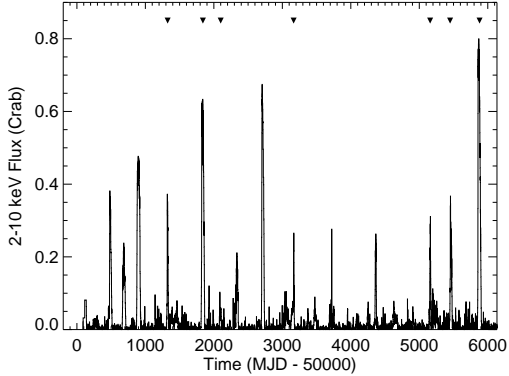}
\caption{\textit{RXTE}/ASM and MAXI 2--10 keV light curve of Aql
X$-$1. Light curve after MJD 55500 is MAXI/2--10 keV light curve.
Triangles indicate the seven outbursts we analyzed.
\label{fig:pcalc}}
\end{figure}

As shown in Figure~\ref{fig:pcalc}, the combined \textit{RXTE}/ASM
and MAXI 2--10 keV light curve of Aql X$-$1 indicates 16 outbursts in
total in a period of 16 years, of which the \textit{RXTE}
observations covered the hard X-ray state during the rising phase of
seven outbursts, which occurred in 1999, 2000, 2001, 2004, 2009, 2010
and 2011, respectively, as indicated by the triangles in the plot.

\begin{figure*}
\plotone{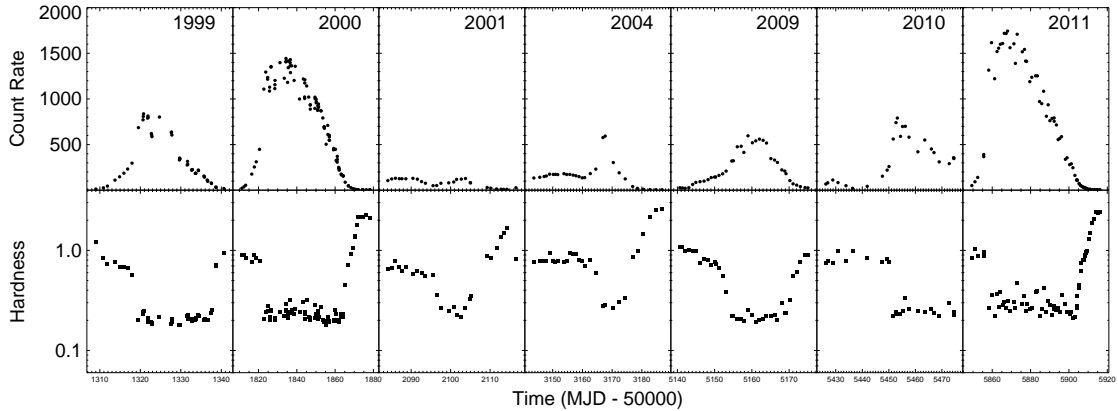}
\caption{Evolution of the PCA count rates and the X-ray hardness
  ratio of the seven outbursts for which hard states during rising
  part of the outburst were covered by the \textit{RXTE}. The PCA
  Std2 2--20 keV light curves are plotted in the upper panels while
  the corresponding 9--20 keV/2--9 keV hardness ratios are plotted in
  the lower panels. The year when the outbursts began are indicated
  in the upper right corner of each panel.\label{fig:lchr}}
\end{figure*}

The PCA 2--20 keV count rate and the HR during the seven outbursts
are shown in Figure~\ref{fig:lchr}. The peak count rate ranged from
$\sim$200 (the 2001 outburst) to $\sim$1800 (the 2011 outburst) $\rm
counts~s^{-1}~PCU^{-1}$, spanning nearly one order of magnitude in
the \textit{RXTE} observations. For the 1999, 2000, 2004, 2010 and
2011 outbursts, the hard-to-soft state transitions are clearly shown
by the abrupt decrease of the HR from $\sim$1.0 to $\sim$0.2. For the
2001 outburst, the hard-to-soft state transition was identified in
\citet{yu_hard--soft_2007}, which was made based on HEXTE(15--250
keV)/PCA(2--9 keV) hardness ratio. So we used the previous
identification. The time of the hard-to-soft state transition
corresponds to the observation 60054-02-04-00 on MJD 52095. It is
also worth noting that the 2001 outburst was peculiar since it was
not a Fast Rise with Exponential Decay (FRED) outburst as others. In
addition, the hard-to-soft state transition in the 2001 outburst was
also peculiar since it occurred when the luminosity declined
\citet{yu_hard--soft_2007}. For the 2009 outburst, we identified the
time corresponding to the hard-to-soft state transition based on
source timing properties. This is based on the fact that for other
atoll sources, such as 4U 1608$-$52, the state transition is
accompanied by the transition of properties of the power spectra (see
for example, Figure 7 of \citealt{van_straaten_atoll_2003}). In Aql
X$-$1 we saw such a transition in the power spectra from the
observation 94076-01-04-04 taken on MJD 55152.2 to the observation
94076-01-04-05 taken on MJD 55153.1, hence we can conclude that the
source left the hard state between these two observations.


\subsection{The Evolution of the Power Spectra}

\begin{figure*}
\plotone{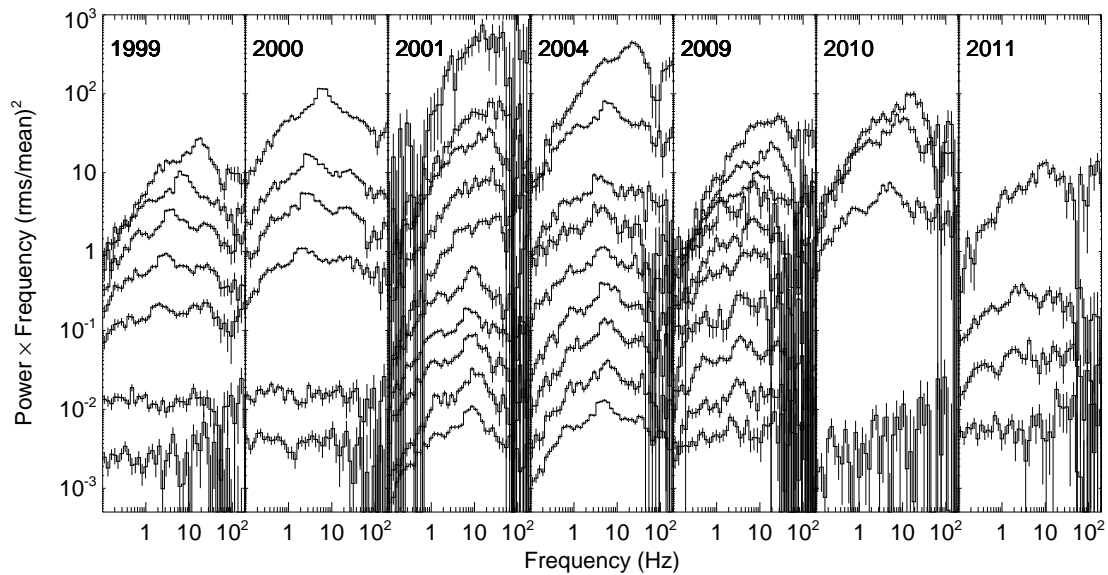} 
\caption{The evolution of the Fourier power spectra in the seven
  X-ray outbursts of Aql X$-$1. The power has been shifted vertically
  by a factor proportional to the time separation between
  observations. The years when these outbursts started are indicated
  in the upper-left corner of each panel. For the 2001, 2004, 2009,
  and 2010 outbursts, the power spectra seen in several observations
  obtained at the beginning of the outbursts are not shown due to
  limited space in the figures. \label{fig:pdsevo}}
\end{figure*}

Figure~\ref{fig:pdsevo} shows the evolution of the power spectra in
the rising phase of the seven outbursts during which the LFQPOs and
the band-limited noise components were seen. The most obvious
component below 100 Hz is the LFQPO component of which the frequency
evolved gradually from several Hz to tens of Hz. The LFQPO component
also corresponds to the maxima in the $\nu P_{\nu}$ vs. Frequency
plot below 100 Hz. We have found that the LFQPOs had the following
properties:

\begin{enumerate}
  \item In all except the 2001 outburst, the frequency increased with
	time before the state transition. This indicates that the
	characteristic timescale corresponding to the LFQPO decreased
	with time during the rising phase of the outbursts.

 \item In all except the 2001 outburst, the frequency appeared to
increase more rapidly just before the hard-to-soft state transition
than during the earlier stages. This is best-illustrated in the 1999
outburst. Notice that the power spectra are plotted in the
logarithmic scale and the power spectra are shifted by multiplying a
factor proportional to the time separation between the observations.
\end{enumerate}

\begin{figure}
\plotone{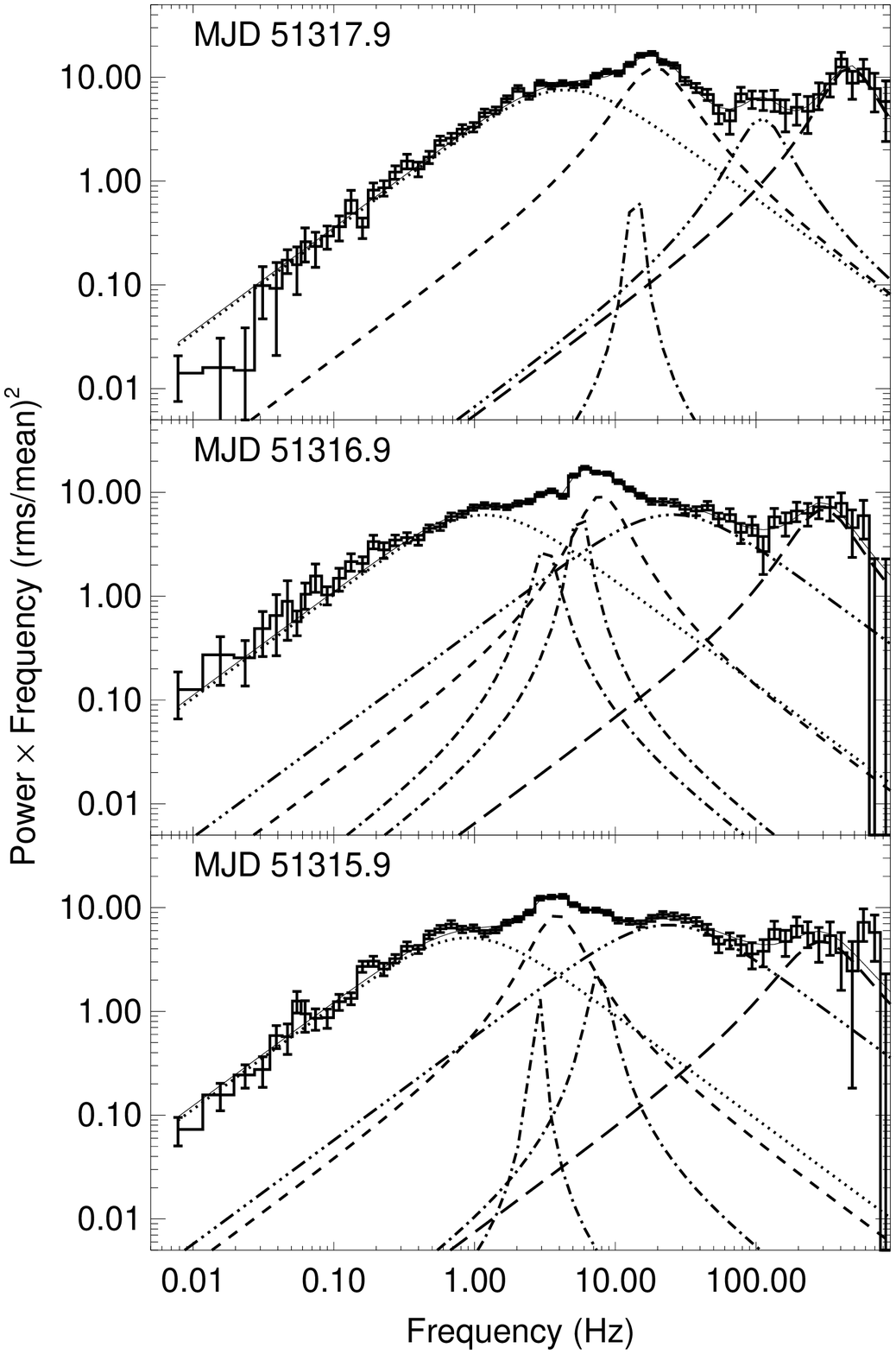}
\caption{The power spectra corresponding to three observations during
  the 1999 outburst of Aql X$-$1. The data are plotted with solid
  lines and the best-fit models are over-plotted. We identified those
  components and plotted the $L_b$, $L_{lf}$, the harmonics of
  $L_{lf}$, $L_l$ and $L_u$ in dotted, dash-dot, dashed,
  dash-dot-dot-dot, and long dashed lines, respectively.
  \label{fig:pdsfit1}}
\end{figure}

To quantitatively analyze the frequency evolution of the LFQPO and
other noise components, we fit the power spectra with
multi-Lorentzians following the previous approach
\citep{belloni_unified_2002}. By doing so the power spectra can be
decomposed into several Lorentzian components empirically
\citep[e.g.,][]{belloni_unified_2002,van_straaten_multi-lorentzian_2002,van_straaten_atoll_2003,
klein-wolt_identification_2008}, allowing us to study frequency
correlations among the characteristic frequencies. Some examples of
our model fits can be seen in Figure~\ref{fig:pdsfit1}, in which we
show the power spectra with the best-fit models corresponding to
three observations performed during the 1999 outburst. The components
with the lowest frequency can be identified as the noise component,
$L_b$, as defined in previous studies. Two distinct components, a
narrow component at a lower frequency and a prominent broader
component with a higher frequency, were found in the frequency range
from several Hz to tens of Hz. The broader and stronger components
are the LFQPOs. In more than half of the observations, the narrow
feature is not required in the power spectral fit. The components
with higher frequencies are identified as $L_l$ and $L_u$, following
the conventions used in \citet{belloni_unified_2002}.

\begin{figure}
\epsscale{1.0}
\plotone{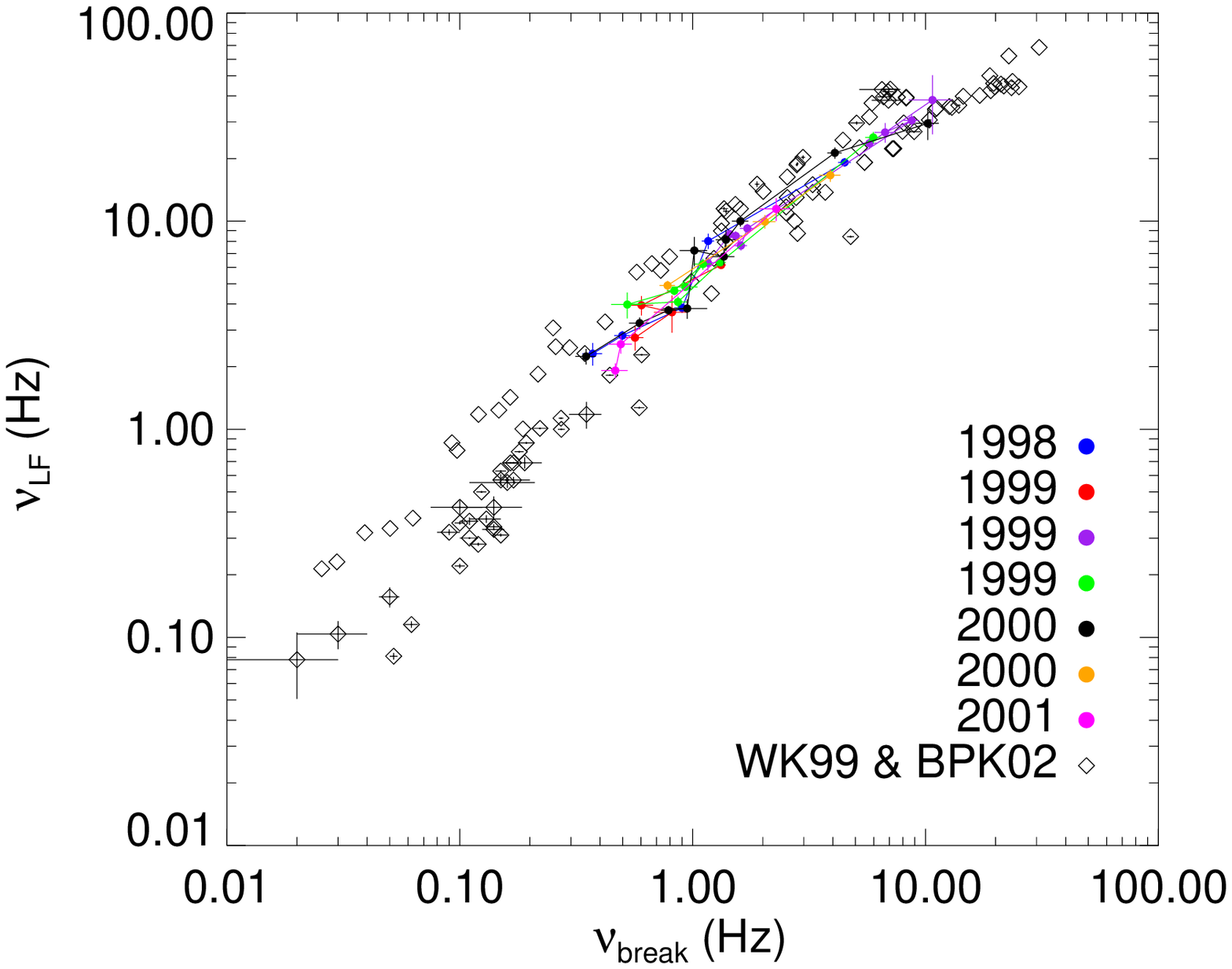}
\caption{The correlation between $\nu_{LF}$ and $\nu_b$ in Aql X$-$1.
The open diamonds represent the correlation in other black
hole/neutron star SXTs reported in \citet{wijnands_broadband_1999}
and \citet{belloni_unified_2002}, and the solid circles represent our
results on Aql X$-$1. The correlation in Aql X$-$1 agrees well with
the correlations established in other SXTs. \label{fig:corr}}
\end{figure}

It has been found that the characteristic frequencies of broad noise
and narrow QPO components are correlated
\citep[e.g.][]{wijnands_broadband_1999,psaltis_correlations_1999,belloni_unified_2002}.
To compare with the results obtained from other sources and to study
how these characteristic frequencies are correlated during single
outbursts in the \textit{RXTE} era, we plotted the LFQPO
characteristic frequency $\nu_{LF}$ vs. break frequency $\nu_b$,
along with the data shown in the WK99 relation
\citep{wijnands_broadband_1999} and the BPK02 relation
\citep{belloni_unified_2002} in Figure~\ref{fig:corr}. The
correlation of $\nu_{LF}$ and $\nu_b$ in Aql X$-$1 is consistent with
the correlation in other sources as expected. 

\subsection{Maximum LFQPO Frequency and the Evolution Pattern}
\label{sec:evo}

\begin{figure}
\plotone{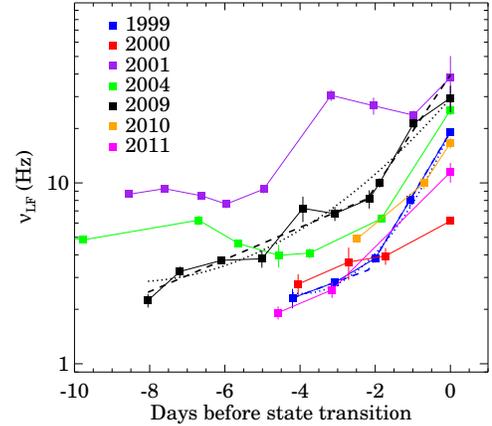}
\caption{The evolution of the frequency of the LFQPO $\nu_{LF}$
during the outbursts of Aql X$-$1. Zero day corresponds to the time
of the last observation before the corresponding soft-to-hard state
transition. The evolution of $\nu_{LF}$ in the 1999 (blue) and the
2009 (black) outburst was fit with an exponential quadratic
polynomial and a broken exponential model, and the best-fit models
are plotted in dotted and dashed lines,
respectively.\label{fig:qpoevo}}
\end{figure}

The evolution of the characteristic frequency of the LFQPOs in seven
outbursts is shown in Figure~\ref{fig:qpoevo}. First we examined the
maximum LFQPO frequency the source can reach in the hard states; this
frequency may correspond to a certain radius or timescale in the
accretion flow when the state transition occurs. As seen from
Figures~\ref{fig:qpoevo}, the maximum LFQPO frequency varies among
outbursts, from $6\pm0~\rm Hz$ in the 2000 outburst to $38\pm12~\rm
Hz$ in the 2001 outburst, although the uncertainty of the latter is
quite large. In the 2009 outburst, the maximum frequency was
$30\pm5~\rm Hz$. We took the weighted average of the maximum LFQPO
frequency of the 1999 and the 2001 outburst, i.e., $\sim31~\rm Hz$,
as the estimate of the maximum LFQPO frequency seen in Aql X$-$1 in
the following discussion.

The evolution pattern of $\nu_{LF}$ was quite similar among the
outbursts. In the following analysis, for the 2004 outburst, we
concentrated on the last four observations in the hard state before
the hard-to-soft transition, since earlier observations corresponded
to a special flare in the light curve, as can be seen from
Figure~\ref{fig:lchr}. For the peculiar 2001 outburst, the frequency
was around 10 Hz at the beginning, then increased rapidly to around
30 Hz at about 6 days before the state transition, and then decreased
to around 24 Hz at about just one day before the transition, and
finally turned to around 40 Hz at the time of the state transition.
For all the other outbursts, the frequency of the LFQPO increased
with time consistently during the rising phase of the outbursts.

\begin{figure}
\plotone{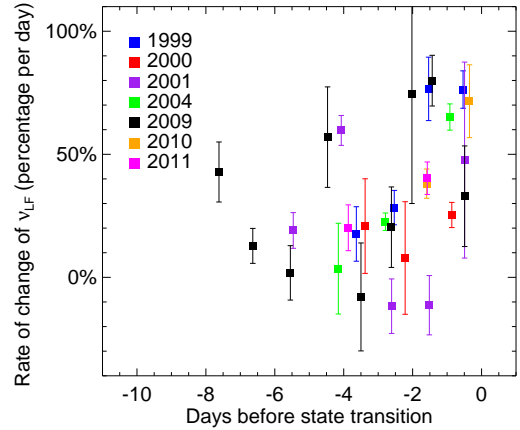}
\caption{The fractional rate-of-change of
  $\nu_{LF}$ vs. time during the seven outbursts of Aql X$-$1, as
  shown in different colors. Day zero corresponds to the time of the
  last observation just before the hard-to-soft state transition. The
  fractional rate-of-change is the rate-of-change of the frequencies
  divided by the frequencies themselves.\label{fig:qpoder}}
\end{figure}

In our model-independent approach, we estimated the rate-of-change of
the LFQPO frequency, following $\dot{\nu}=(\nu_2-\nu_1)/(t_2-t_1)$,
where $\nu$ is the LFQPO frequency, $t$ is the time, and the
subscripts 1 and 2 denote two consecutive observations. Then we
calculated the fractional rate-of-change of the frequency
$2\dot{\nu}/(\nu_1+\nu_2)$, corresponding to the time at
$(t_2+t_1)/2$. Figure~\ref{fig:qpoder} shows the time evolution of
the fractional rate-of-change. The fractional rate-of-change
increased from 10\% day$^{-1}$ at the beginning of the outburst, to
$\sim$100\% day$^{-1}$ just before the state transition.

The pattern of the increase of the LFQPO frequency deviates from a
single exponential pattern, which would follow a straight line if
plotted in logarithmic scales, and the fractional rate-of-change
would be constant with time. We tried two phenomenological models to
fit the pattern. The first one is an exponential quadratic
polynomial, i.e. $\nu=e^{at^2+bt+c}$. This function is continuous and
the fractional rate-of-change increases with time when the parameter
{\it a} is positive; the second one is of a broken exponential
function, i.e., the frequency first increases with a constant
e-folding timescale $t_1$, then afterwards the frequency increases
with another constant e-folding timescale $t_2$. We fit the models to
the data of the 1999 outburst and the 2009 outburst which have the
most number of data points available (which is also more than the
number of model parameters). The dashed and the dotted lines show
the best-fit broken exponential and exponential quadratic polynomial
models, respectively. Notice that the time at zero corresponds to
the time when the last observation of the source in the hard state
was taken.

For the 2009 outburst, the $\chi^2$ were 8.76 with 6 Degrees of
Freedom (DOF) and 34.25 with 7 DOF for the broken exponential
function and exponential quadratic polynomial function, respectively,
indicating that the former model describes the pattern of the
frequency evolution better. The result from the best-fit broken
exponential model showed that the frequency increased with an e-fold
timescale of $5.13\pm0.17$ and $1.31\pm0.17$ days before and after
the moment at $2.80\pm0.22$ days before the state transition,
respectively. For the 1999 outburst, the model fit with an
exponential quadratic polynomial gave a good fit with $\chi^2=1.89$
with 2 DOF, while the fit with the broken exponential model yielded a
$\chi^2=0.11$ with 1 DOF, indicating we were lacking statistics to
constrain the later model. The trend in both the 2009 outbursts
suggests that the frequency increased with time at different speed in
the early and in the late stage of the hard state during the rising
phase of an outburst, further supporting the idea that the transition
is associated with non-stationary accretion during these outbursts.

\subsection{LFQPO Frequency and Its Dependence on the X-ray Flux}
\label{sec:frevsflux}

\begin{figure}
\plotone{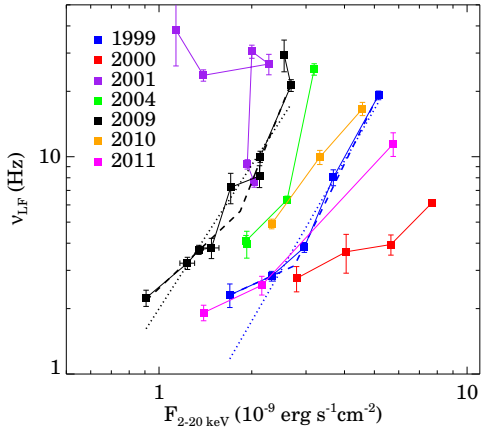}
\caption{The $\nu_{LF}$ vs. 2--20 keV source flux relation of Aql
X$-$1. Notice that the same QPO frequency corresponds to different
X-ray fluxes, and vice versa. We fit the data corresponding to the
1999 (blue) and the 2009 (black) outburst with a power-law model and
a broken power-law model, and the best-fit models are plotted as
dotted and dashed lines, respectively.\label{fig:frevsflux}}
 \end{figure}

In Figure~\ref{fig:frevsflux} we show how the LFQPO frequency evolved
with the 2--20 keV flux. During the rising phase of an outburst, the
frequency in general increased with the X-ray flux, except for the
peculiar 2001 outburst. The frequency-flux relation for each outburst
deviates from a straight line on a log-log plot, showing that it does
not follow a simple power-law relation. The power-law model was only
acceptable for the 2000 outburst, with $\chi^2=0.59$ and 1 DOF, while
for other outbursts the model was rejected. The model parameters in
the fits with the power-law model are tabulated in
Table~\ref{tab:fit}. We fit the data with a broken power-law model
instead. Such a model consists of four free parameters: the
normalization, the break frequency, and the power-law indices before
and after the break. We fit the model to the data corresponding to
the 1999 and the 2009 outburst which have more than four data points. 

\begin{deluxetable}{rrrrr}
\tabletypesize{\scriptsize}
\tablecaption{Power-law model fit to the frequency vs. flux
relation\label{tab:fit}}
\tablewidth{0pt}
\tablehead{
\colhead{Year} & \colhead{$\alpha$} & \colhead{N} &
\colhead{$\chi^2$} & \colhead{DOF}}
\startdata
1999 & $2.45\pm0.08$ &$0.47\pm0.05$ & 36.80 & 3	\\
2000 & $0.88\pm0.12$ & $0.27\pm0.07$ & 3.48 & 2	\\
2004 & $2.50\pm0.22$ & $1.04\pm0.23$ & 98.50 & 2	\\ 
2009 & $2.20\pm0.09$ & $2.23\pm0.13$ & 39.36 & 7	\\
2010 & $1.81\pm0.12$ & $0.79\pm0.11$ & 0.59 & 1	\\
2011 & $1.26\pm0.12$ & $0.45\pm0.05$ & 4.89 & 1	
\enddata
\tablecomments{The parameters of best-fit power-law model of
 frequency vs. 2--20 keV flux: $\nu_{10}=Nf_{3}^{\alpha}$, where
 $\nu_{10}=\nu_{LF}/10~\rm Hz$ and $f_3$ is flux in
 $3\times10^{-9}~\rm erg~s^{-1}cm^{-2}$.}
\end{deluxetable}


For the 1999 outburst, the broken power-law model yielded a
$\chi^2=1.36$ with 1 DOF while the power-law model gave a
$\chi^2=36.80$ with 3 DOF, showing a preference for the broken
power-law model. In that model, the break occurred at $\sim 2~\rm
Hz$. The index before the break was not well constrained, giving
$0.64\pm0.43$, while the index after the break was well-constrained
as $2.89\pm0.12$. Fitting with the broken power-law model yielded a
$\chi^2=12.35$ with 5 DOF, while the power-law model fit gave a
$\chi^2=39.36$ with 7 DOF. The break frequency occurred at $\sim7~\rm
Hz$. The power-law indices were $1.34\pm0.22$ and $3.58\pm0.34$
before and after the break, respectively. The best-fit power-law and
broken power-law models are plotted as dashed and dotted lines,
respectively, in Figure~\ref{fig:frevsflux}. 

Among different outbursts, the power spectra show that the source can
reach significantly different characteristic QPO frequencies at
similar X-ray fluxes. For instance, as shown in
Figure~\ref{fig:frevsflux}, at a source flux of $2\times10^{-9}~\rm
erg~cm^{-2}~s^{-1}$, the frequency of the LFQPO was at around 2 Hz
for the 1999 and the 2011 outburst, but at around 30 Hz for the 2001
outburst. It is interesting to notice that for a certain
characteristic frequency, the X-ray flux is usually higher in the
outbursts in the years of 1999, 2000 and 2011, which were associated
with higher state transition fluxes, while the X-ray flux
corresponding to the same QPO frequency was lower in the outbursts in
the years of 2001, 2004, 2009 and 2010, which were associated with
lower state transition fluxes. There is another anti-correlation
between the maximum LFQPO frequency and the state transition flux (or
the outburst peak X-ray luminosity), with the Spearman correlation
coefficient of -0.96 at a significance level of $4.5\times10^{-4}$.
This is shown in the upper-right corner of Fig.~\ref{fig:frevsflux}.
Actually the maximum frequency of the LFQPO was reached during the
2001 and the 2009 outburst, which was associated with the lowest
state transition fluxes.

\section{Conclusion and Discussion}

We have analyzed the observations obtained during the rising phase of
seven outbursts of the neutron star soft X-ray transient Aql X$-$1
with the \textit{RXTE}/PCA when the source was in the hard state. As
expected, in the power spectra, the characteristic break frequencies
of the band-limited noise component and the frequency of the LFQPOs
were correlated, consistent with the overall correlation established
in a large sample of black hole and neutron star X-ray binaries
previously found.

The frequency evolution of LFQPOs is shown in
Figure~\ref{fig:pdsevo}. Both the characteristic frequency of the
band-limited noise and the frequency of the LFQPOs increased with the
PCA/2--20 keV count rate before the hard-to-soft state transitions,
except for the peculiar 2001 outburst. We found that the maximum
frequency of the LFQPOs in Aql X$-$1 can reach 31~{\rm Hz}. We
measured the rate-of-change of the LFQPO frequency during the rising
phase of the outbursts and found that the frequencies increased by
around one tenth per day in the early stage of the outbursts to
nearly one hundred percent per day nearly two days before the
hard-to-soft state transition. In the 2009 outburst, for which we
have the best data available, we found that the frequency of the
LFQPOs actually increased with time with an e-folding timescale of
$5.13\pm0.17$ and $1.31\pm0.17$ days in the early rising phase and in
the later rising phase, respectively.

In Figure~\ref{fig:corr} we also show that the frequency correlation
between band-limited noise and the LFQPOs holds down to the timescale
of one to two days, since the \textit{RXTE} observations were taken
roughly every other day. It would be interesting to investigate the
correlation on shorter time scales with future X-ray missions to test
if such a correlation holds on shorter time scales. In addition, we
found that the characteristic frequency vs. flux relation is
significantly different among the seven outbursts. Higher
characteristic frequency is reached by the source at similar X-ray
flux levels in the outbursts with lower state transition and outburst
peak fluxes. In individual outbursts, the relationship between the
frequency and the source flux deviated from a simple power-law form.
The frequency depended on the source flux. For the 2009 outburst, If
we model it with a broken power-law function, we found that the
power-law indices were $1.34\pm0.22$ and $3.58\pm0.34$ before and
after the break, respectively. The observed multiple relation between
the LFQPO frequency and X-ray flux and the fast variation of the
frequency would potentially put strong constraints on the
Lense-Thirring precession model of the LFQPO proposed by
\citep{ingram2009}, if quantitative prediction of the relation
between the transitional radius, the mass accretion rate and the
Lense-Thirring precession frequency can be delivered. 

\subsection{Maximum LFQPO Frequency: Indicator of the Mass of
the Compact Object ?}

In this work, we found that the maximum frequency of the LFQPOs in
Aql X$-$1 observed with the \textit{RXTE} was $\sim$ 31 Hz. In a
similar study of the black hole transient GX 339$-$4 we found that
the maximum frequency of the LFQPOs was around 5 Hz, which is about
six times lower than that of Aql X$-$1. \cite{ozel_mass_2012} showed
that the mass of the neutron stars in the recycled pulsars has a mean
value of $1.48~\rm M_{\odot}$ with a dispersion of $0.2~\rm
M_{\odot}$. We know that the black hole binary GX 339$-$4 has a mass
function of $5.8~\rm M_{\odot}$ \citep{hynes_dynamical_2003}, so we
do not know the exact mass ratio. If we take its mass as a typical
value of 10 solar masses in black hole X-ray binaries, the mass ratio
between the compact stars in the two systems seems consistent with
the ratio between their maximum frequencies of the LFQPOs. This
suggests that the maximum frequency of the LFQPOs may be an indicator
of the mass of the compact star. By observing the maximum frequency
of the LFQPOs during outbursts, we may be able to determine the
nature of the compact star in a new transient LMXBs empirically.

\subsection{Tracing the Inner Disk Radius with LFQPO ?}

Using the observations we analysed, we can investigate whether it is
possible to trace the evolution of the truncation radius using the
LFQPO based on current disk truncation models and to constrain these
models with observations. It has been believed that the frequency of
one of the twin kHz QPOs might be associated with the Keplerian
frequency at the inner disk edge
\citep[e.g.,][]{miller_sonic-point_1998,stella_lense-thirring_1998,osherovich_kilohertz_1999}.
Therefore we can probably associate the frequency of the LFQPOs with
the inner disk edge (or the truncation radius in some popular models)
based on the empirical frequency correlation between the kHz QPOs and
the LFQPOs. The empirical correlations found in the \textit{RXTE}
data suggest that for the lower kHz QPO at frequency $\nu_1$, the
frequency of the LFQPO is $ \nu_{lf} \approx (42\pm3~\rm Hz)(\nu_1
/500 ~\rm Hz)^{0.95 \pm 0.16}$ \citep{psaltis_correlations_1999}. The
empirical relation between the lower and the upper kHz QPO is $\nu_1
= (724\pm3)\left( \nu_2/1000~\rm Hz \right )^{1.9\pm0.1}~\rm Hz$
\citep{psaltis_beat-frequency_1998}. This yields the relation between
the LFQPO and the upper kHz QPO as $\nu_{LF} \propto
\nu_2^{1.85\pm0.21}$.

In the magnetospheric model
\citep{ghosh_accretion_1977,ghosh_accretion_1979}, the inner edge of
the disk is where the ram pressure is balanced by the magnetic
pressure. \citet{ghosh_accretion_1979} studied the structure of a
geometrically thin disk around a magnetized neutron star and found
that the disk is truncated at $r_0\approx C r_A$, where $r_A$ is the
Alfv\'{e}n radius, i.e., $r_A =\mu^{4/7} (2GM)^{-1/7}
\dot{M}^{-2/7}$, here $\mu=BR^3$ is the magnetic dipole moment and
$R$ is the neutron star radius, and $C$ is a parameter which does not
affect configuration a lot. The magnetospheric model therefore
predicts that the truncation radius $r_{tr} \propto \mu^{4/7}
\dot{M}^{-2/7}$.

In the disk evaporation model, the heat generated in the corona
formed above the disk is conducted down to the lower layer of the
corona and radiated away. If the density of the lower layer is too
low to efficiently radiate away the heat, the thin disk would be
heated and then evaporated, leading to a steady mass flow from the
disk to the corona. In a stationary state, the outer disk would be
truncated at a radius $r_{tr}$ where the evaporation is balanced by
the mass inflow from the outer disk. This truncation radius lies at
$r_{tr} \propto \dot{M}^{-1/1.17}$. Further development of the model
included the effect of the viscosity and the accretion disk magnetic
field \citep{taam_disk_2012,liu_constraints_2013}, which gives the
truncation radius $r_{tr} \propto \dot{M}^{-0.886} \alpha^{0.07}
\beta^{4.61}$, where $\alpha$ is the viscous parameter and $\beta$ is
the ratio of gas pressure to total pressure. The truncation radius
depends only weakly on the viscosity.

It is worth noting that the parallel-track phenomenon of the LFQPO
shown in Fig.~8, which is similar to that seen for the kHz QPOs,
actually suggests that mass accretion rate is not the only parameter
that determines the QPO frequency. However, let's consider the
simplest case where mass accretion rate is the only parameter in the
accretion system. As we know, both disk truncation models predict
that the truncation radius is a function of the mass accretion rate,
i.e., $r_{tr} =c \dot{M}^\delta$, where $c$ is the normalization
coefficient. Then we get the relation between the LFQPO frequency and
the mass accretion rate as $\nu_{LF} \propto
\dot{M}^{1.43\delta}=\dot{M}^{\epsilon_1}$ if taking the lower kHz
QPO as the orbital frequency at the inner disk edge and as $\nu_{LF}
\propto \dot{M} ^{2.71\delta} =\dot{M}^{\epsilon_2}$ if taking the
upper kHz QPO frequency as the Keplerian frequency at the inner disk
edge. For the former case, the index $\epsilon$ is $1.22\pm0.20$ in
the evaporation model and $0.57\pm0.10$ in the magnetospheric model;
while in the later case which corresponds to the upper kHz QPO
assumption, the index $\epsilon$ is $2.31\pm0.26$ in the evaporation
model and $0.77\pm0.10$ in the magnetospheric model. In the above
estimates we have assumed that the bolometric correction factor was a
constant during those observations and that the bolometric luminosity
was proportional to the mass accretion rate (i.e., the radiation
efficiency did not change a lot).

In Section~\ref{sec:frevsflux} we showed that for the well-sampled
2009 outburst, the frequency vs. the X-ray flux relation was best-fit
by a broken power-law model with indices of $1.34\pm0.22$ and
$3.58\pm0.34$ before and after the trend break. Both of the values
are larger than that predicted by the magnetospheric model. As for
the disk evaporation model, we found that for the case corresponding
to the lower kHz QPO assumption, during the early rising phase of the
outburst, the observed index before the trend break is consistent
with the model, while after the break the observed index is much
larger than that predicted by the disk evaporation model. This might
indicate that the disk evaporation model is consistent with the
observed evolution of the inner disk edge at the early rising phase
of those outbursts but not later on. For the case corresponding to
the upper kHz QPO assumption, the disk evaporation model
overestimates the index in the early stage of the outburst but
underestimates the index in the later stage. This suggests that at
least additional physics should be considered in the current disk
evaporation model. We also realised that additional constraints can
be put on the evaporation model as well from the so-called
`parallel-track' phenomenon of the LFQPO. In the improved evaporation
model investigated in \citet{taam_disk_2012}, the truncation radius,
$r_{tr}$, depends on the magnetic field in the accretion disk as
$r_{tr} \propto \dot{M}^{-0.886} \alpha^{0.07} \beta^{4.61}$. In
order to explain the observations showing that lower characteristic
frequency can be observed under similar mass accretion rates in the
outbursts with a higher state transition luminosity, the disk
evaporation theory with the consideration of the magnetic field has
to be consistent with the fact the brighter an outburst is, the lower
the magnetic pressure in the accretion disk should be, due to the
fact that there is empirical correlation between the luminosity of
the hard-to-soft transition and the peak luminosity of the outburst
or the following soft state
\citep{yu_correlation_2004,yu2007,yu_state_2009}.


\acknowledgments
We would like to thank the anonymous referee for useful comments
which have improved the work a lot.
The authors would like to thank Michiel van der Klis, Diego
Altamirano and Ronald Taam for useful discussions and comments. This
work was supported in part by the National Natural Science Foundation
of China under grant No. 11333005, 11073043, and 11350110498, by
Strategic Priority Research Programme ``The Emergence of Cosmological
Structures'' under Grant No. XDB09000000 and the XTP project under
Grant No. XDA04060604, by the Shanghai Astronomical Observatory Key
Project and by the Chinese Academy of Sciences Fellowship for Young
International Scientists Grant.
This research has made use of data and software provided by the
High Energy Astrophysics Science Archive Research Center (HEASARC),
which is a service of the Astrophysics Science Division at NASA/GSFC
and the High Energy Astrophysics Division of the Smithsonian
Astrophysical Observatory.



{\it Facilities:} \facility{\textit{RXTE} (PCA)}

\end{document}